\documentclass[aps,prd,showpacs,nofootinbib,floats,floatfix,preprintnumbers,groupedaddress,twocolumn]{revtex4}

\usepackage{bm}
\usepackage{latexsym}
\usepackage{dcolumn}
\usepackage{amsmath,amsfonts,amssymb}
\usepackage{graphicx,epsfig}

\usepackage{fancyhdr}

\begin{document}
\title{Noncomutativity in near horizon symmetries in gravity}
\author{Bibhas Ranjan Majhi}
\email{bibhas.majhi@iitg.ernet.in}
\affiliation{Department of Physics, Indian Institute of Technology Guwahati,\\
 Guwahati 781039, Assam, India}

\date{\today}

\begin{abstract}
We have a new observation that near horizon symmetry generators, corresponding to diffeomorphisms which leave the horizon structure invariant,  satisfy noncommutative Heisenberg algebra. The results are valid for any null surfaces (which has Rindler structure in the near null surface limit) and in any spacetime dimensions. Using Sugawara construction technique the central charge is identified. It is shown that the horizon entropy is consistent with the standard form of Cardy formula. Therefore we feel that the noncommutative algebra might leads to quantum mechanics of horizon  and also can probe into the microscopic description of entropy.
\end{abstract}

\maketitle


\section{Introduction}
   Semiclassical treatment of gravity leads to the fact that black holes behave like a ordinary thermodynamical objects \cite{Bekenstein:1973ur,Hawking:1974sw} -- the horizon attributes entropy which, in general relativity (GR), is equal to one quarter to the horizon area. Although it was originally thought that only black holes are the candidates which incorporates temperature and entropy, later it was found that any null surface (not necessarily the solution of Einstein's equations) has similar properties \cite{Parattu:2013gwa}. This observation is the main basis for the thought of the gravity as an emergent phenomenon \cite{Padmanabhan:2009vy}. The idea is locally there is a Rindler observer (the uniformly accelerated frame on the Minkowski spacetime) which attributes temperature and entropy on the Rindler horizon. So the accelerated observer plays the same role of the static observer in a black hole spacetime. Therefore, it is usually thought that such an universality of thermodynamical behaviour of null surfaces, among which Rindler one is most simpler, can give some hints of quantum nature of gravity.
    
    The microscopic description of entropy can be put forwarded by using asymptotic symmetries of spacetime and Cardy formula \cite{Cardy:1986ie} in sense that in the original ($1+1$) dimensional conformal field theory (CFT) this has been obtain by the microstates counting. The concept for black holes was first introduced in \cite{Brown:1986nw} which was later on, extended by Carlip \cite{Carlip:1999cy} to any Killing horizon by using near horizon symmetries. The underlying idea of this method is to identify the relevant diffeomorphisms which preserve certain symmetries and using them calculate the Fourier modes of the Noether charge and the bracket among them. This leads to Virasoro algebra with central extension. Finding the central charge one calculates the entropy with the use of Cardy formula.  
    
    Recently, it has been noticed that the symmetry preserving conditions only for the horizon (not the angular part of the metric) is sufficient to obtain such result \cite{Majhi:2012tf}. Roughly speaking the relevant diffeomorphims preserves the near horizon structure and hence we have a reduced theory which contains a subset of all the generators. In a more physical sense, among all the degrees of freedom some of them, which were originally gauge degrees of freedom, raised to true degrees of freedom due to the particular choice of symmetry and lead to horizon entropy.   
    
    In this paper we adopt that same boundary condition and define two different Noether charges corresponding to two components of diffeomorphism vector. It is found that the bracket among them for GR is trivial; i.e. there is no central extension. But a linear combination of them; i.e. changing the basis, we find much more enriched structure -- {\it the new modes of charges obey noncommutative Heisenberg algebra}. This is quite surprising and may enlighten the microstates responsible for entropy. Also following the Sugawara construction \cite{book1}, the central charge is identified. Using this the Cardy formula, consistent with the horizon entropy, is derived. This is done by relating the zero mode of Noether charge to entropy.  
    
    Let us now discuss more about our results. We show that there are three sets of possibilities in the values of the brackets among the new set of Fourier modes of charges. In all cases the ``Hamiltonian'' and ``timelike'' variable commute with all canonical pairs and also these canonical pairs obey usual Heisenberg algebra among them. But the two sets of canonical variables are either both or any one of them are noncommutative . Such a similar algebra was first observed in String theory \cite{Seiberg:1999vs} where spacetime was itself noncommutative. Here we find that in the context of gravity -- the near horizon symmetries, conserving the horizon structure, can lead to similar algebra. This is completely new and quite unexpected. 
    
    Very recently in three dimensional case the usual Heisenberg algebra (not noncommutative one) has been obtained \cite{Afshar:2016wfy,Afshar:2016uax}. In this analysis the boundary condition is completely different from us as it has been imposed on the time and angular sector of the metric. Here we are not doing the same. Our boundary condition does not deal with angular parts; more precisely, as it will be cleared later, we impose this on the retarded time and radial sector. Moreover, the earlier one is restricted to ($1+2$) dimensions while the present one is much more general as it is applicable to any dimensions. In addition the analysis is done on a Rindler metric which is a null surface, observed from a local frame and the Noether charge we use is {\it off-shell} one i.e. derived without using Einstein's equations (See \cite{book2} for details). Therefore the results are applicable beyond the black hole solutions.
    
    The significance of the results, we think, is very important. First of all, it tells that the imposed boundary condition raises some of the gauge degrees of freedom to true degrees of freedom which contribute to entropy. Moreover, the new observation -- noncommutativity in generators -- can lead us to think the horizon in a quantum mechanical way. We shall discuss this again at the end.
Lets now go into the calculation part.      

\section{Setup: near horizon symmetry and charges}
   As we said that locally one always find a null surface and a most simplest one can be represented by Rindler metric. For simplicity and clarity, here the calculation will be done based on this line element. The same for a general null metric will be given elsewhere. 
   The $D$ spacetime dimensional Rindler metric in retarded (outgoing) Eddington-Finkelstein coordinates is given by
\begin{equation}
ds^2=-2ax du^2-2dudx+dx_{\perp}^2~.
\label{metric}
\end{equation} 
 In the above $a$ is the acceleration of the Rindler observer and $x_{\perp}$ are the ($D-2$) transverse coordinates.  The Killing horizon is located at $x=0$.
Imposition of the conditions $\pounds_{\xi}g_{xx}=0=\pounds_{\xi}g_{xu}$, where $\pounds_{\xi}$ is the Lie derivative along the vector $\xi^a$, we find the components of the vector as \cite{Majhi:2012tf}
\begin{equation}
\xi^u = F(u,x_{\perp}); \,\,\,\ \xi^x=-x\partial_u F(u,x_{\perp})~,
\label{xi}
\end{equation}
where $F$ is an arbitrary function, independent of $x$. For the above components, it is easy to check that $\pounds_{\xi}g_{uu}\sim\mathcal{O}(x)$ which vanishes near the horizon. This implies that the obtained diffeomorphism vector (\ref{xi}) is Killing vector in the near horizon region only for the ($u-x$) sector; not for the whole metric. Moreover, it does not change the horizon structure; i.e. the location of it does not change. Hence we call them as horizon structure invariant boundary conditions.
It must be pointed out that the same diffeomorphims, given in (\ref{xi}), play important role in various aspects of gravity. The entropy of horizon was calculated in the context of Virasoro algebra and Cardy formula by using the Noether charge corresponding to York-Gibbons-Hawking surface term (See also \cite{Chakraborty:2016dwb} for a general null surface). Also they have significant importance in hydrodynamics in gravity \cite{Eling:2012xa} and in membrane paradigm and horizon BMS symmetry \cite{Eling:2016xlx}. Here in this paper we explore more about them.

The Noether charge, calculated on the horizon, is defined as
\begin{equation}
Q[\xi] = \frac{1}{2}\int_{\mathcal{H}}d\Sigma_{ab}J^{ab}[\xi]~,
\label{Q}
\end{equation} 
where $d\Sigma_{ab} = d^{D-2}x_{\perp}\sqrt{\sigma}(l_ak_b-l_bk_a)$ and $J^{ab}[\xi]=(1/16\pi G)(\nabla^a\xi^b-\nabla^b\xi^a)$ for GR.  $\mathcal{H}$ stands for horizon. Here $l^a$ is null vector near the horizon while $k^a$ is another null vector which satisfies $l^ak_a=-1$. The bracket among the charges is given in \cite{Majhi:2011ws}:
\begin{eqnarray}
&&[Q_1,Q_2] =\pounds_{\xi_1}Q[\xi_2] - \pounds_{\xi_2}Q[\xi_1]
\nonumber
\\
&&=\int_{\mathcal{H}}d\Sigma_{ab}\Big(\xi_2^aJ^b[\xi_1]-\xi_1^aJ^b[\xi_2]\Big)~,
\label{bracket}
\end{eqnarray}
where the Noether current is $J^a[\xi]=\nabla_bJ^{ab}[\xi]$.

  Before going to the main purpose, let us discuss about the other existing charges in literature. It is interesting to note that our bracket (\ref{bracket}) exactly matches with Carlip's result (First paper of \cite{Carlip:1999cy}) for General relativity (GR). Whereas this differs from Silva's value \cite{Silva:2002jq} by a factor which is proportional to $\nabla^a(\xi_1^c\nabla_c\xi_2^b - \xi_2^c\nabla_c\xi_1^b) - (a\leftrightarrow b)$ upto some total derivative term. A detailed discussion can be followed from the Appendix A of \cite{Silva:2002jq}. As it has been explained there by examples that this extra term does change the value of the zero mode of the charge but not the value of central charge. On the other hand, there is another charge by Barnich and Brandt \cite{Barnich:2001jy} which has been recently used in \cite{Donnay:2016ejv} to study the extended symmetries at the black hole horizon. It can be noted that such expression differs from the Silva's one by a factor which is proportional to $(\nabla^c\xi_1^b+\nabla^b\xi_1^c)(\nabla^a\xi_{2c}+\nabla_c\xi_2^a) - (a\leftrightarrow b)$. This can be checked by comparing the Eq. (8) of \cite{Silva:2002jq} and Eq. (6.25) of \cite{Barnich:2001jy}. Therefore the present one (\ref{bracket}) differs by the above two terms from \cite{Barnich:2001jy}. Of course it would be interesting to calculate these for the vectors (\ref{xi}), discussed here. But for the present purpose we shall concentrate on (\ref{bracket}).

   For the metric (\ref{metric}), we choose $l^a=(1,0,\bf{0})$ (this is the Killing vector as the metric is independent of $u$) while  $k^a$ can be taken to be the normal to the $u=$ constant surface. Therefore, the components of $k^a$ are $(0,1,\bf{0})$. Consequently, the covariant components are $l_a=(-2ax,-1,\bf{0})$ and $k_a = (-1,0,\bf{0})$.  Now notice that the diffeomorphism vector $\xi^a$ has two components. Here our aim is to calculate the charges and the brackets corresponding to two different components. For that we define two vectors: one of which has only $x$ component, given by $\xi^x$; while other one has only $u$ component, given by $\xi^u$.  Therefore let us denote the charges (\ref{Q}) and the brackets (\ref{bracket}) for respective vectors, as
\begin{equation}
Q^{\pm} \equiv Q[\xi^\pm] = \frac{1}{2}\int_{\mathcal{H}}  d\Sigma_{ab}J^{ab}[\xi^\pm]~,
\label{Qpm}
\end{equation} 
and
\begin{eqnarray}
&&[Q_1^\pm,Q_2^\pm]=\int_{\mathcal{H}}d\Sigma_{ab}\Big(\xi_2^aJ^b[\xi_1^\pm]-\xi_1^aJ^b[\xi_2^\pm]\Big)~,
\label{QQpm}
\\
&&[Q_1^+,Q_2^-] = \int_{\mathcal{H}}d\Sigma_{ab}\Big(\xi_2^aJ^b[\xi_1^+]-\xi_1^aJ^b[\xi_2^-]\Big)~,
\label{Q+Q-}
\end{eqnarray}
where $\xi^+ = (0,\xi^x,\bf{0})$, $\xi^- = (\xi^u,0,\bf{0})$ and $\xi^a = \xi^+ + \xi^- = (\xi^u,\xi^x,\bf{0})$.
For the present metric, after a laborious calculations one obtains:
\begin{eqnarray}
&&Q^{+} = -\frac{1}{16\pi G}\int_{\mathcal{H}}d^{D-2}x_{\perp} \partial_uF;
\nonumber
\\
&&Q^- = -\frac{1}{16\pi G}\int_{\mathcal{H}}d^{D-2}x_{\perp} (\partial_uF - 2aF)~;
\label{QF}
\\
&&[Q_1^+,Q_2^+] = \frac{1}{16\pi G}\int_{\mathcal{H}}d^{D-2}x_{\perp}[F_2\partial^2_uF_1 -(1\leftrightarrow 2)]~;
\nonumber
\\
&&[Q_1^-,Q_2^-] =\frac{1}{16\pi G} \int_{\mathcal{H}}d^{D-2}x_{\perp}[(F_2\partial^2_uF_1-2aF_2\partial_uF_1)
\nonumber
\\
&&-(1\leftrightarrow 2)]~;
\nonumber
\\
&&[Q_1^+,Q_2^-] = \frac{1}{16\pi G}\int_{\mathcal{H}}d^{D-2}x_{\perp}[F_2\partial^2_uF_1-F_1\partial_u^2F_2
\nonumber
\\
&&+2aF_1\partial_uF_2]~.
\label{QQF}
\end{eqnarray}
Next we shall calculate the Fourier modes of the above quantities and provide the main results of this letter.

\section{Results}
   Take the Fourier modes of $F$ as $F_m = (1/a)\exp[im(au+p.x_{\perp})]$ where $m$ and $p$ include all positive or negative integers.  Here both in prefactor and exponential the parameter is chosen to be acceleration $a$ as the retarded time has periodicity $0$ to $2\pi/a$ so that $F_m$ becomes a periodic function of $u$ and to ensure that the diffemorphism vector satisfies $i\{\xi_m,\xi_n\}^a = (m-n)\xi_{m+n}^a$. The same has been pointed in earlier works \cite{Carlip:1999cy,Silva:2002jq}. This is found to be consistent with the black hole thermodynamics from gravitational action \cite{Gibbons:1976ue}. The similar compactification on $u$ was also used recently \cite{Afshar:2015wjm} in the context of Virasoro algebra and black holes in three spacetime dimensions. The idea for the above choice can also be elaborated the following way. First of all $F_m$ is function of both $u$ and $x_\perp$ and also it is that Fourier modes of function $F$. Therefore $F_m$ must satisfies the two equations $\partial^2F_m/\partial u^2+a^2m^2F_m=0$ and $\partial^2F_m/\partial x_{\perp}^2+(pm)^2F_m=0$ so that it should be at least linear combination of sin and cos functions and in that case it is also periodic. This is precisely the choice of the paper which satisfies these equations and the periodicity is coming automatically as $F_m$ is the Fourier modes. The more deeper reason can be the following. There is a planer symmetry in the perpendicular direction and hence in this direction one can argue for plain wave expansion. On the other hand, since there is an equilibrium horizon one can has periodicity in time with the Euclidean sector in mind. 
   
Then from (\ref{QF}) and (\ref{QQF}) one can easily find the corresponding Fourier modes:
\begin{eqnarray}
&&Q_m^+ = 0; \,\,\,\ Q_m^- = \frac{A}{8\pi G}\delta_{m,0}~;
\label{Q_m}
\\
&&[Q_m^+,Q_n^+]=0; \,\,\,\ [Q_m^-,Q_n^-] = -im\frac{A}{4\pi G}\delta_{m+n,0}; 
\nonumber
\\
&& [Q_m^+,Q_n^-] = -im \frac{A}{8\pi G}\delta_{m+n,0}~,
\label{QQm}
\end{eqnarray}
where $A$ is the transverse area of the horizon.
Note that the positive modes of the charge; i.e. for $\xi^+$, always vanish. Therefore we define the entropy of the horizon as
\begin{equation}
S = 2\pi  Q_0^- = \frac{A}{4G}~;
\label{entropy}
\end{equation} 
i.e. only the $\xi^-$ contributes.
Having obtained the above results, we are now in a position to show the existence of non-triviality within them which was claimed in the introduction.  

   Define new modes of charges as different combinations of the above charges; i.e. express the near horizon symmetry generators in new basis. Consider the following combinations:
\begin{eqnarray}
&&P_0 = Q_0^+ + Q_0^-; \,\,\ P_m = \mathcal{A} Q_{-m}^+ + \mathcal{B}Q_{-m}^- \,\,\,\ {\textrm{(for $m\neq 0$)}}~;
\nonumber
\\
&&X_m = \mathcal{C}Q_{m}^+ + \mathcal{D}Q_{m}^-~,
\label{new}
\end{eqnarray}
where the coefficients $\mathcal{A}, \mathcal{B}$ and $\mathcal{C}, \mathcal{D}$ are all non-zero. Below we present three interesting choices of these coefficients, compatible in every respect.

\noindent   
{\bf{Case I:}} For the choice $\mathcal{A} = -1/(mC_0)\pm(1/C_0)\sqrt{1+1/m^2} - (1/m\pm\sqrt{1/m^2+1})$, $\mathcal{B} = 1/m\pm\sqrt{1/m^2+1}$, $\mathcal{C}=- (1+1/C_0)$ and $\mathcal{D}=1$ where $C_0 = A/4\pi G$, the non-zero brackets are
\begin{eqnarray}
&&[X_m,X_n]=\frac{i}{2}(m-n)\delta_{m+n,0} = [P_m,P_n]~;
\nonumber
\\
&&[X_m,P_n]=i\delta_{m,n}~.
\label{Case1}
\end{eqnarray}

\noindent
{\bf{Case II:}} Choose $\mathcal{A}=-2/m$, $\mathcal{B}=2/m$, $\mathcal{C}=- (1+1/C_0)$ and $\mathcal{D}=1$. Then the nontrivial ones are
\begin{eqnarray}
&&[X_m,X_n]=\frac{i}{2}(m-n)\delta_{m+n,0}; \,\,\,\ [P_m,P_n]=0~;
\nonumber
\\
&&[X_m,P_n]=i\delta_{m,n}~.
\label{Case2}
\end{eqnarray}

\noindent
{\bf{Case III:}} If we have the following choice of coefficients: $\mathcal{A}=2/(C_0m)-m/2$, $\mathcal{B}=m/2$, $\mathcal{C}=-\mathcal{D}=1$, then one obtains
\begin{eqnarray}
&&[X_m,X_n]=0; \,\,\,\ [P_m,P_n]=\frac{i}{2}(m-n)\delta_{m+n,0}~;
\nonumber
\\
&&[X_m,P_n]=i\delta_{m,n}~,
\label{Case3}
\end{eqnarray}
while the others vanish.
Note that in all cases we find different notions of non-commutative Heisenberg algebra. These are the main results of the paper. Of course, the present ones are not exactly similar to the earlier obtained non-commutative algebra \cite{Seiberg:1999vs} in the context of String theory. Here the non-commutativity exists only for $m=-n$ while the other one shows such structure for all $m,n$ with $m\neq n$. Hence we call the present result as ``restricted'' non-commutative algebra. Moreover, in the original non-commutative algebra (known as Snyder algebra) is between the space coordinates. Here we can not say that our new variables ($X_m$ or $P_m$) are exactly those. Actually we do not know the meaning of these at this stage. So the whole similarity is just structure wise; not in one to one sense. Therefore at this position we can say in gravity case we have some variables which behaves similar to Snyder variables at the algebra level; not at the physical level.
     
      Having obtained the above interesting algebra, now our aim is to observe the entropy (\ref{entropy}) in the microstate counting manner. Following Sugawara construction \cite{book1}, we define new generators as
\begin{equation}
L_m^{\pm} =\frac{1}{2C_0} \sum_p Q_{m-p}^{\pm}Q_p^{\pm} + imQ_m^{\pm}~.
\label{L}
\end{equation}   
Then it is easy to verify that $L_m^-$ satisfies the Virasoro algebra:
\begin{equation}
i[L_m^-,L_n^-] = (m-n)L_{m+n}^-+m^3C_0\delta_{m+n,0}~,
\label{Virasoro}
\end{equation}  
with the central charge $C=12C_0$.
Also from (\ref{L}) one can identify that $Q_0^- = \sqrt{2C_0L_0^-}=\sqrt{(CL_0^-)/6}$. Therefore, the entropy (\ref{entropy}) can be expressed in the form of Cardy formula \cite{Cardy:1986ie}:
\begin{equation}
S = 2\pi \sqrt{\frac{CL_0^-}{6}}~.
\label{Cardy}
\end{equation}
Also note that $P_0$ can be related to the ``surface'' Hamiltonian of the system. The idea comes from a very well known result exists in literature \cite{Majhi:2013jpk}. The surface part of the gravitational action, calculated on the horizon, yields $A_{sur} = -t TS$ where $t$ is the timelike coordinate and $T=a/(2\pi)$ is the temperature. Then the semi-classical surface Hamiltonian is given by $H_{sur} = -\partial A_{sur}/\partial t=TS$. Now since here $Q_0^+=0$ and $Q_0^-=A/(8\pi G)$, we define our surface Hamiltonian as $H_{sur} = aP_0$. Hence we can say, following the algebra given in Eqs. (\ref{Case1}), (\ref{Case2}) and (\ref{Case3}), that $H_{sur}$ commutes with all canonical conjugate variables $X_m$ and $P_m$.  Then by the arguments of \cite{Afshar:2016wfy}, we can say that horizon must carry ``soft hair'' \cite{Hawking:2016msc} which, for the present case is soft noncommutative Heisenberg hair, and also it does not contribute to the entropy. 

\section{Summary and outlook}
   In this letter, we took the Rindler metric as a null one, seen by a local observer, whose horizon is the null surface. Imposing the simple condition that the horizon structure remains invariant the relevant class of diffeomorphims have been identified. Then we showed that in a certain basis the modes of the corresponding charges satisfy ``restricted'' non-commutative algebra. In this sense the near horizon structure is much more richer that what we usually expect. The observation is completely new and very interesting as the non-commutativity can lead to quantum structure of horizon.
   
   We used the word ``surprisingly'' as no body has looked at the charges for two components of diffeomorphisms and hence such features were  unexplored. Also there is no obvious way, without calculating, to tell that there is some non-triviality. Earlier people usually calculated them for the ``total'' vector, consists of two or more components and it is not at all indicative that such analysis can lead to non-commutativity. Here we took the components as two different vectors and calculated brackets for them (Eqs. (\ref{Q_m}), (\ref{QQm})). This is new as no body has looked in this manner and it revels a new feature: there is a non-commutativity which is more vivid in the new basis. The similar has been done recently in \cite{Afshar:2016wfy,Afshar:2016uax} but that is too restrictive as it is confined within the three dimensions and the algebra is usual Heisenberg one. Whereas, this one is valid in any dimensions and algebra is different for the earlier analysis. Therefore, although (\ref{Case1}), (\ref{Case2}) and (\ref{Case3}) may be obvious from the results (\ref{Q_m}) and (\ref{QQm}), but later ones are not. 
   
   It must be pointed out that the present diffeomorphism algebra in ($D-1$) coordinates $u, x_\perp$, is much larger than a Witt-algebra (usually the Witt-algebra is the one dimensional diffeomorphism algebra, whereas F is a function of $D-1$ coordinates and so it should form a $D-1$-dimensional diffeomorphism algebra). Still the present results reduces to similar to one dimensional. The reason is as follows. In the usual ($1+1$) dimensional conformal field theory the charges are defined as the integration over one coordinate with the integrand is relevant component of energy-momentum tensor. Therefore it reduces to one dimensional structure. Whereas, for black hole metric the charge is defined by the integration of $J^{ab}$ over the all transverse coordinates. So it is evident that all the transverse coordinates are integrated out and so we are left with only two coordinates: one timelike and another one is spacelike (Here these are $u$ and $x$, respectively). Moreover our calculation is near the horizon where the theory is effectively two dimensional with the effective metric is given by the ($u-x$) sector and also the function $F$ is a function of $u,x_\perp$ one of which is integrated out. Now it is well known that such a theory is conformal one and hence we are getting one dimensional result even if our original theory is $D$ dimensonal.
      
   Let us now explain why it is being demanded that the non-commutativity can lead to quantum nature of the horizon. One well known fact about the usual non-commutativity in space-space or momentum-momentum is they lead to quantisation of space or phase-space. This is because, $[x^a,x^b]=i\Theta^{ab}$ leads to a quantity, a linear combination of the square of these variables, which is quantised. In the present analysis we are also getting similar non-commutative variables and in the same manner these may lead to quantum structure of horizon. Here we are not demanding the quantization of any specific horizon quantity; rather a hope arises that this might give some clue towards it. We agree that it is not sure at this stage which one is quantized as the meaning of the non-commutative variables are not clear at all. But people are looking at these possibilities (See for Refs. \cite{Afshar:2016wfy,Afshar:2016uax}) along the similar line and hence it is worth to investigate some of these issues more closely. The present paper is exactly in this direction. 
      
   Moreover we showed that zero mode of the charge is related to horizon entropy which can be expressed in the form of Cardy formula. The implication of it is these horizon structure invariant diffemorphisms corresponds to some microstates, which are yet to identify, lead to the entropy of the null surface.  Additionally, it may be worth to mention that the imposed conditions on the metric coefficients raises some of the gauge degrees of freedom to true ones which contribute to the entropy. Although, we are still not a place to pin point ``the'' states; nevertheless, we feel such an analysis can illuminate towards the microstates counting as it was already discussed that the entropy, with the help of the present charges, can be expressed in the form of the Cardy formula.
   
   It would be interesting to understand the meaning of two copies of charges ($Q^+$ and $Q^-$) corresponding to respective components of a diffeomorphism vector which keeps the horizon structure invariant. In this context let us point out that for $\xi^+$ and $\xi^-$ the changes in the metric coefficients are as follows: $\delta g_{uu}^+ \sim \mathcal{O}(x)$, $\delta g_{ux}^+ = \partial_uF$ and $\delta g_{xx}^+ = 0$ while $\delta g_{uu}^- \sim \mathcal{O} (x)$, $\delta g_{ux}^- = -\partial_u F$ and $\delta g_{xx}^- = 0$, respectively. This shows that for individual components at least the location of the horizon does not change. The situation is much more interesting in null coordinates ($u,v$) with $dv=du+dx/ax$. In this coordinates, the changes in metric coefficients are: $\delta g_{uu}^+ \sim \mathcal{O}(x)$, $\delta g_{uv}^+ \sim \mathcal{O}(x)$ and $\delta g_{vv}^+ = 0$ while $\delta g_{uu}^- \sim \mathcal{O} (x)$, $\delta g_{uv}^-  \sim \mathcal{O}(x)$ and $\delta g_{vv}^- = 0$. This implies that metric remains ``invariant'' very near to the horizon for $\xi^+$ or $\xi^-$ alone. Moreover, the components of the vectors are $\xi^+ = (0,-(1/2a)F,\bf{0})$ and $\xi^- = (F,F,\bf{0})$; i.e. one is along $v$-direction while the other one is the resultant of both $u$ and $v$ components. What we found here is the algebra between the charges due to diffeomorphisms along these two directions, which is surprisingly noncommutative in nature. But the complete physical meaning is still not known and also what are the consequences of it is beyond the scope of the paper.  Nevertheless, the results are quite interesting and may play a very significant role in exploring the quantum nature of horizon.
   
   An interesting point can also be noted from the above discussion. Here we found the diffeomorphism vectors (\ref{xi}) by using a particular type of ``fall-off'' condition in the Eddington-Finkelstein coordinates. But when the components of them are treated as separate vectors ($\xi^+$ and $\xi^-$), then they individually do not preserve the asymptotic boundary condition in this gauge. Therefore a natural question comes into the picture: Are the charges $Q^{\pm}$ (see Eq. (\ref{QF})) correspond to any well-defined asymptotic symmetry? From the above discussion one can argue that the answer is ``Yes''. The reason is as follows. We already mentioned that our asymptotic condition is the invariance of horizon structure. The meaning of it is the dffeomorphism vectors should change the metric such that the ($u-x$) sector of the metric remains unchanged on the horizon. Therefore the relevant metric coefficients can change by minimum of order $x$ and hence the diffeomorphisms are Killing vectors on the horizon. As a result the location of the horizon does not change. This was our original motivation for choosing the total $\xi^a$ in Eddington-Finkelstein coordinates. This is a ``weaker'' condition than one usually takes in a particular gauge. The above analysis shows that interestingly the vectors $\xi^{\pm}$ preserves this asymptotic boundary condition in ``null-null'' coordinates and so they are Killing vectors on the horizon for the ($u-v$) sector of the metric. Also the location of horizon remains same in null-null coordinates. This can be explicitly checked by taking the norm of $\xi^{\pm}$ is equal to zero. Since the charges are defined as scalar quantity one obtains the same value in ``null-null'' case also. Hence, in this sense, the charges obtained here correspond to our present asymptotic symmetry in null-null coordinates.  
   
   As a final remark, we want to mention that the results are may be very general and goes beyond the Rindler form as null surface. It is now in under investigation if the similar structure  also follows for a general null metric. The analysis is not straight forward and will be reported separately with much more details \cite{Krishna} since the calculation involves some technicalities as all metric coefficients depend on all coordinates. Nevertheless, the present analysis incorporates all the stationary spacetimes with a null surface as even the Kerr like form reduces to Rindler one in the near horizon limit. This is OK since the whole discussion is valid near the null surface. Therefore the results, obtained here, is very much {\it local} in nature. Moreover, since these are based on the specific choice of our boundary conditions, they are the features of some particular class of observers who are consistent with these conditions; more precisely the algebra is observer dependent. Also it is independent of spacetime dimensions and the approach is general enough to be extended to any theory of gravity.  In summary, the present results may show a window to search the true quantum theory of a horizon (more generally of a null surface), which yet to be found in a concrete sense. 
\vskip 9mm
\section*{Acknowledgments}
I wish to thank Daniel Grumiller and T. Padmanabhan for critical comments on the first draft. I also thank Krishnakanta Bhattacharya for checking all my calculations.
The research of the author is supported by a START-UP RESEARCH GRANT (No. SG/PHY/P/BRM/01) from Indian Institute of Technology Guwahati, India.



\end{document}